\documentstyle[prb,aps,epsf,multicol]{revtex}
\begin{document}
\title{Length distribution of single walled carbon nanotubes determined by ac atomic force 
microscopy
}
\author{Richard Piner and Rodney S. Ruoff$^{a)}$
}
\address{
Department of Mechanical Engineering\\
Northwestern University\\
2145 Sheridan Road\\
Evanston, IL 60208-3111\\
}
\date{\today}
\maketitle

\begin{abstract}
A simple method to disperse individual single walled carbon nanotubes ( SWCNT ) on an 
atomically flat substrate is presented.  Proper tuning of ac modes of atomic force microscopes 
(AFM) is discussed.  This is needed to discriminate between individual nanotubes and very small 
bundles.  The distribution of lengths of the nanotubes measured by these methods is reported.
\end{abstract}
\ \\

a) Author to whom correspondence should be addressed;
electronic mail: r-ruoff@northwestern.edu
\begin{multicols}{2}
Since the discovery of carbon nanotubes, there has been much interest in their mechanical and 
electrical properties.  However their extreme small size makes measurement of their physical 
properties very challenging.  The smallest diameter tubes are single walled carbon nanotubes 
(SWCNT).  Current methods of producing SWCNTs do not produce tubes of uniform length. 
One would like to know just what the distribution of lengths is in a given SWCNT sample. 
However, the answer to this question is not as straight forward as one might expect.
 
There are two major challenges to measuring the length of SWCNTs.  First, the tubes are always 
bundled together.  These ``ropes'' make it impossible to determine the length of an individual 
tube.  The second problem is related to the perfect surface of the tube and it's low adhesion to 
suitable substrates.  Any attempt to scan the tubes by atomic force microscope (AFM) on a 
substrate such as mica, usually results in displacing the tubes.  In this paper, we will present 
solutions to both these problems and a length distribution for one sample of SWCNT material. 

We used the following procedure to disperse individual SWCNTs on an atomically flat substrate. 
Our starting material is a SWCNT sample obtained from Tubes@Rice.  These tubes were grown 
by the laser ablation method\cite{rinzler98} and were delivered in a toluene suspension.  After trying many 
different methods to disperse these tubes, we found that the simplest was also the best.  We take 
a few milliliters of the obtained suspension and put it directly into dimethylformamide (DMF), at 
a dilution ratio of 1:100.  Then the vial (50 ml glass, teflon cap, pre-cleaned; from Wheaton, 
``Clean-Pa'', Wheaton No. 217847) is sonicated in a small ultrasonic bath (Crest, model 175HT, 
capacity about 1 liter) for about 4 hours.  The resulting suspension is then further diluted 1:100 in 
DMF and sonicated for 4 more hours.  This procedure is then repeated a third time.  The final 
suspension is then further sonicated for another 4 hours.  Thus the total dilution is 1:1,000,000 
and the total sonication was typically ~16h.  The result is a clear suspension with about 50\% 
percent of the tubes completely separated as individual tubes.  The suspension seems to be 
completely stable, with no precipitation after 3 months.  We found that the simplest way to get 
the correct dilution was to test each suspension as we went along by AFM.  Our procedure was 
to test after each sonication step.  (Thus, if the final solution is a little too dilute, a few drops of 
the second suspension can be added to achieve the desired concentration.) 

To image the SWCNTs, they were deposited on a freshly cleaved mica surface.  To do this, two 
pieces of mica were cleaved.  Then a large drop of suspension was placed on one piece and the 
second piece was then placed on top of the first.  This results in a uniform film of DMF/tubes 
sandwiched between two pieces of mica.  This alters the surface tension and drying dynamics of 
the DMF such that the SWCNTs remain uniformly distributed across the mica surfaces as the 
DMF evaporates.  The mica sandwich is allowed to air dry over night. 

AFM imaging of these tubes was also difficult.  Since the SWCNTs can be so easily moved from 
the substrate, standard contact mode imaging proved to be almost impossible.  However, in 
recent years, new AC methods of AFM have become available.  The methods include non-
contact and intermittent-contact modes.  We used two different AFMs to perform these 
experiments, one was a model CP, the second a CP Research, both from Thermomicroscopes. 
The cantilevers were unmounted non-contact ultralevers, also from Thermomicroscopes.  These 
AC methods are very sensitive to the tuning of the instrument.  The design of the detector heads 
has changed quite a bit between these two instruments.  So, we believe that our observations on 
tuning technique are generally true for AC modes and not an instrumental artifact. 

When tuning the frequency, after laser and detector have been aligned, the frequency is scanned 
and the amplitude of oscillation is plotted.  We have found that it is often the case that there are 
several peaks in the resulting spectrum.  We have also found that if this happens, the cantilever 
can be taken out and shifted a few microns in its chip carrier.  When remounted, the frequency 
response can be quite different. Two such spectra are show in {Fig. 1.} It is important to keep 
adjusting the mounting of the cantilever chip until the spectrum shows a single peak in 
amplitude.  Once this is done, it is possible to reliably set the correct frequency for scanning 
SWCNTs.  We have found that higher frequencies seem to work better.  Typically we work with 
cantilevers with resonances around 200 kHz.  We will start with the frequency set about 500 Hz 
below resonance.  Once the tip has engaged, the feedback loop should be stable at a gain factor 
of 0.5 or above.  If there appears to be a feedback oscillation, we lower the frequency by 20 Hz 
and try again.  When the correct frequency is achieved the feedback is stable at gains above 0.5, 
and we have been able to go as high as 2.0 before the feedback becomes unstable.  Typical scan 
speeds are 1 Hz for a 4 micron image.  When the above parameters are used, very stable images 
of SWCNTs can be acquired.  Repeated scans show no movement of the tubes.

\begin{figure}[t,b,p]
\vspace{.1in}
\hspace{.25in}
{\epsfxsize=2.5in
\epsffile{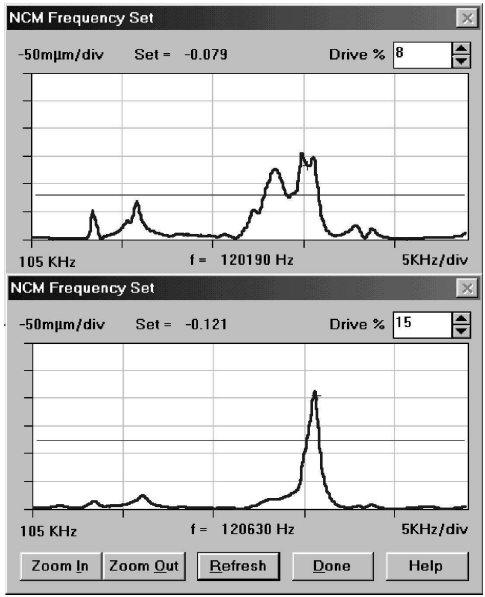} 
\vspace{.25in}
}
\caption{Example of frequency tuning for intermittent contact mode imaging.  This figure shows 
two screen captures for cantilever tuning.  The top spectrum shows multiple peaks, while the 
lower shows one strong peak.  Both spectra are from the same cantilever, the only difference is a 
slight shift in mounting position of the chip in its carrier.
}
\end{figure}

In AFM, when there is a problem, there is always the question, is it the 
instrument, or the sample?  We have found that there is a simple test sample, which can be 
used to test for proper operation of the AFM.  A glass microscope slide is cut into a piece small 
enough to fit into the AFM.  It is then cleaned with Windex$^{TM}$, and vigorous rubbing with a 
clean paper towel.  If the AFM is operating well, and the tip is sharp, the glass surface will be 
decorated with small circles (Fig.2).  These circles have been seen in both AC modes and lateral 
force mode (LFM).  This has proven to be an inexpensive reproducible test substrate for AFM 
operation. If the circles are well defined, good resolution images of SWCNTs will result, if the 
SWCNT sample is well prepared. 

\begin{figure}[t,b,p]
\hspace{.0in}
{\epsfxsize=3.0in
\epsffile{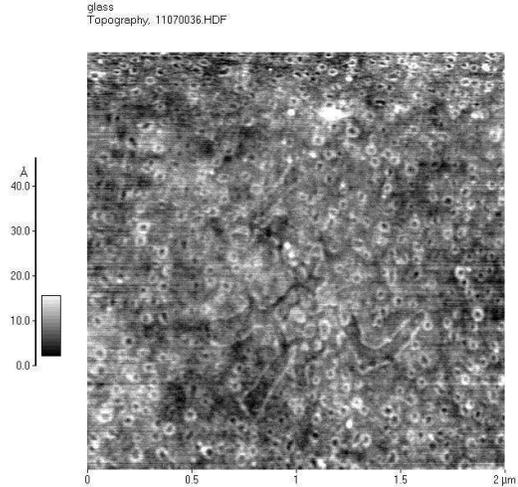} 
\vspace{.2in}
}
\caption{Intermittent contact image of a glass microscope slide.  This image can be used as a test to 
determine correct operation of intermittent contact and the quality of the tip.  Scan size is 2 
microns.
}
\end{figure}

The reason that such precise tuning is required is so that individual SWCNTs can be 
distinguished from small bundles made up of just a few tubes.  Once we have done this tuning 
operation correctly, we find that the apparent diameter of the tubes, as measured by the obtained 
height, is only about 0.8 nm.\cite{postma00} This is well below the nominal 
value of 1.2 to 1.4 nm that one would expect for these samples.\cite{smalley96}  However, we are scanning 
in air, on mica.  Mica is extremely hydrophilic, and there is a fair amount of water adsorbed on 
the mica surface. The dynamics of AC AFM methods with water are still not well understood.  
This could be the origin of the smaller than expected diameter measurements of the tubes.  In any 
case, these tubes are clearly distinguished from bundles, both of which appear in all of our 
images.  (As a side note, we have noticed that when relative humidity in the lab drops below 
20\%, scanning becomes unstable and contrast reversal over the tubes can be the result.  
However, at humidity between 20 and 40\%, results are very reproducible.)
 
Once we have these images, we used the built-in image measuring software to measure the 
lengths of individual tubes.  Each image was first flattened to eliminate the curvature from the 
bending of the scanning tube.  Then the contrast of the image is reversed as seen in Fig. 3.  In 
reverse contrast, higher objects appear darker.  Individual tubes will be light.  This is easier to 
discern when working on a video monitor.  In order to check that we were measuring the lengths 
of only the smallest diameter objects, a few select images were examined with the instrument's 
built-in height measurement tool.  An example of this is shown in Fig. 4.  By comparing heights with 
contrast, it is possible to 
learn how to distinguish between bundles and individual tubes by just the image contrast.  While 
this introduces some small error, it also speeds data analysis significantly. 

\begin{figure}[t,b,p]
\hspace{0.0in}
{\epsfxsize=3.5in
\epsffile{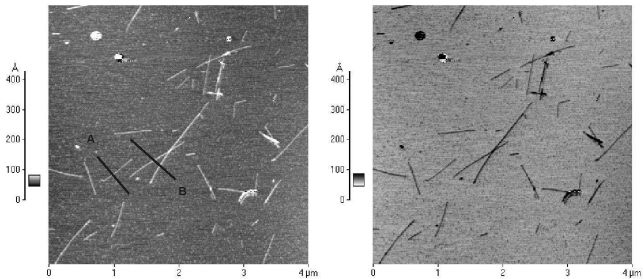} 
}
\vspace{0.1in}
\caption{Reverse contrast image of SWCNTs. This is used to determine which objects to measure.
Scan size is 4 microns.
Lines (A) and (B) indicate cross sections shown in Fig. 4.} 
\end{figure}
\protect{
\begin{figure}[t,b,p]
\hspace{.5in}
{\epsfysize=3.5 in
\vspace{0.25in}
\epsffile{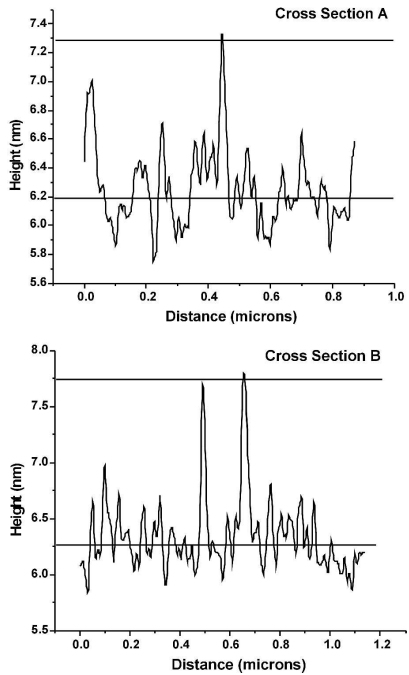}
}
\caption{
Typical cross sections taken from image in Fig. 3. See lines drawn in Fig. 4.
Note that the larger bundles look darker in the previous figure. In this example, diameter of single
tube is 1.0nm, while the two bundles have a diameter of 1.5nm.
}
\end{figure}
}

Using this technique, the lengths of approximately 500 individual tubes were measured.  
The program saves the lengths to a file.  
The recorded lengths from each image are compiled into a single file, and then the histogram of 
lengths is computed.  The resulting histogram is shown in Fig. 5. 

\protect{
\begin{figure}[t,b,p]
\hspace{0.1in}
{\epsfxsize=3.5in
\epsffile{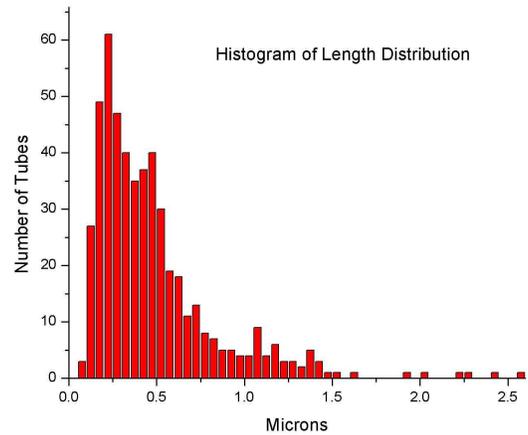} 
\vspace{.25in}
}
\caption{Histogram of tube lengths based on a measurement of the lengths of 500 tubes. The peak 
in the histogram falls at 300 nm; the mean is 480 nm.
}
\end{figure}
}

In summary, we have measured the length distribution of individual carbon nanotubes, with a 
peak in the distribution of 300 nm.  A simple method to prepare high quality samples for 
scanning is repeated dilution and sonication in DMF.  With proper tuning, intermittent contact 
methods of AFM can distinguish between single tubes, and small bundles.

The authors wish to acknowledge (prior) support from the Office of Naval
Research and from a current grant from the NASA Langley Research Center
Computational Materials: Nanotechnology Modeling and Simulation Program,
that have allowed this project to be finished.
\bibliography{lendis}
\end{multicols}
\end{document}